\documentclass[12pt]{article}
\def\aap{\ {A\&A}\ }

\def\aaps{\ {A\&AS}\ }

\def\aj{\ {AJ}\ }

\def\apj{\ {ApJ}\ }
\def\apjl{\ {ApJL}\ }
\def\apjs{\ {ApJS}\ }

\def\araa{\ {ARA\&A}\ }

\def\bain{\ {Bul. Astron. Inst. Neth.}\ }

\def\mnras{\ {MNRAS}\ }

\def\newa{\ {New Astron.}\ }

\def\pasp{\ {PASP}\ }
\def\pasj{\ {Publ. Astr. Soc. Japan}\ }

\usepackage{scicite}
\usepackage{times}
\usepackage{epsfig}

\newcommand{\be}{\begin{eqnarray}}
\newcommand{\ee}{\end{eqnarray}}

\newcommand{\Rsun}{\mbox{${\rm R}_\odot$}}

\newcommand{\Msun}{\mbox{${\rm M}_\odot$}}
\def\apgt{\ {\raise-.5ex\hbox{$\buildrel>\over\sim$}}\ }
\def\aplt{\ {\raise-.5ex\hbox{$\buildrel<\over\sim$}}\ }

\newenvironment{sciabstract}{%
\begin{quote} \bf}
{\end{quote}}

\newcounter{lastnote}

\title{The Origin of OB Runaway Stars}

\date{}

\author{Michiko S.\, Fujii$^{1,2}$, Simon Portegies Zwart$^{1\ast}$  \\
       \\
       \normalsize{$^{1}$ Leiden Observatory, Leiden University,}\\ 
       \normalsize{P.O. Box 9513, 2300 RA Leiden, The Netherlands} \\
       \normalsize{$^{2}$ Graduate School of Science and Engineering} \\
       \normalsize{1-21-35 Korimoto, Kagoshima 890-0065, Japan}\\
       \normalsize{Kagoshima University,} \\
       \normalsize{$^\ast$To whom correspondence should be addressed; E-mail:  spz@strw.leidenuniv.nl}
}

\begin{document}

\maketitle

%
% Abstract
%
\begin{sciabstract}
  About 20\% of all massive stars in the Milky Way have unusually high
  velocities, the origin of which has puzzled astronomers for half a
  century.  We argue that these velocities originate from strong
  gravitational interactions between single stars and binaries in the
  centers of star clusters. The ejecting binary forms naturally during
  the collapse of a young ($\aplt 1$\,Myr) star cluster.  This model
  replicates the key characteristics of OB runaways in our galaxy and
  it explains the $\apgt 100$\,\Msun\, runaway stars around young star
  clusters, e.g. R136 and Westerlund~2.  The high proportion and the
  distributions in mass and velocity of runaways in the Milky Way is
  reproduced if the majority of massive stars are born in dense and
  relatively low-mass ($5000-10000$\,\Msun) clusters.
\end{sciabstract}

Most stars in our galaxy have a relatively low velocity. However,
there is a population of fast moving stars, called OB runaways
\cite{1954ApJ...119..625B}, which have considerably higher space
motions of $>30$\,km/s \cite{1986ApJS...61..419G}. The origin of such
velocities can be attained in two very distinct ways: A runaway can be
launched when its binary companion explodes in a supernova
\cite{1961BAN....15..265B}, or by the ejection via a dynamical
slingshot \cite{1988AJ.....96..222L}. The relative importance of both
mechanisms has remained elusive, mainly because both are associated
with young stellar populations and the high speed of a star is
generally observed long after it has moved away from its birth place.

A massive star can be accelerated effectively by a three-body
dynamical interaction \cite{1988AJ.....96..222L}.  Every hard
interaction eventually results in a collision between two or all three
participating stars, or in the escape of one star and one binary
\cite{1975MNRAS.173..729H}.  The velocity acquired by the ejected star
easily exceeds the escape speed of a star cluster: For a binary with
total mass $M_b$ and semi-major axis $a$, the typical velocity with
which the single star is ejected is $v_{\rm ej}^2 = GM_b/a$.

It is typically the least massive star that is ejected
\cite{1975MNRAS.173..729H}, and both components of the ejecting
(bully) binary are therefore likely to be more massive than the
escaping star.  For higher-mass binaries, the maximum cross section
for ejecting an intruder with high speed shifts to a larger orbital
period, and to a higher binding energy (Supplement \S\,A).  The binary
that most effectively produces massive runaways is relatively wide,
composed of the most massive stars in the cluster and located in the
core.  The gravothermal collapse of a cluster core naturally produces
such a binary \cite{2011arXiv1107.3866T}, which subsequently hardens
by ejecting stars \cite{1991ApJ...383..181M}, until it experiences a
collision or hardens to $\sim 10,000$\,kT \cite{Note_on_kT}, upon
which an encounter causes its ejection from the cluster
\cite{2000ApJ...528L..17P}

A binary with a binding energy of 10 to 10,000~kT is therefore a
natural consequence of the core collapse of a star cluster
\cite{2009PASJ...61..721T}, and leads to the frequent ejection of
other cluster members, until it eventually ejects itself or engages in
a phase of Roche-lobe overflow.  During the hardening process, each
encounter typically drains $\sim 40$\% \cite{1975MNRAS.173..729H} of
the binary's binding energy. In this process it ejects on the order of
$\log(10,000{\rm kT}/10{\rm kT})/\log(1+0.4) \simeq 21$ stars from the
cluster before the binary ejects itself or becomes semi-detached.  The
collapsed cluster core can support only one bully binary at a time and
as a consequence the number of runaways produced does not depend on
cluster mass.

We quantized these results by performing a series of $N$-body
simulations for clusters of 6,300\,\Msun\, to $2.0\times
10^5$\,\Msun\, (Tab.\,S1).  For each simulation, we recorded the
moment of core collapse and the time of ejection, mass and velocity of
the escaping stars.  Each cluster produced a single binary that
ejected $\sim 23.0\pm5.9$ stars with $v_{\rm ej}>30$\,km/s of which
$\sim 5.2\pm1.6$ stars have a mass $m >8$\,\Msun, irrespective of the
mass of the cluster (Supplement\,\S\,B); this number is consistent
with the 6 runaways with $m>8$\,\Msun\, ejected from R136
\cite{2010A&A...519A..33G,2011A&A...530L..14B}.  With a mass of $M
\simeq 60,000$\,\Msun\, and a core radius of $r_{\rm core} \aplt
0.10$\,pc \cite{2009ApJ...707.1347A,2010ARA&A..48..431P} R136
represents the population of young ($\sim 2$\,Myr), dense and massive
star clusters in the local group of galaxies (Fig.\,S2). This star
cluster is ideally suited for comparing with our simulations because
of the excellent quality of the copious observations.

\begin{figure}[htbp]
\begin{center}
  \psfig{figure=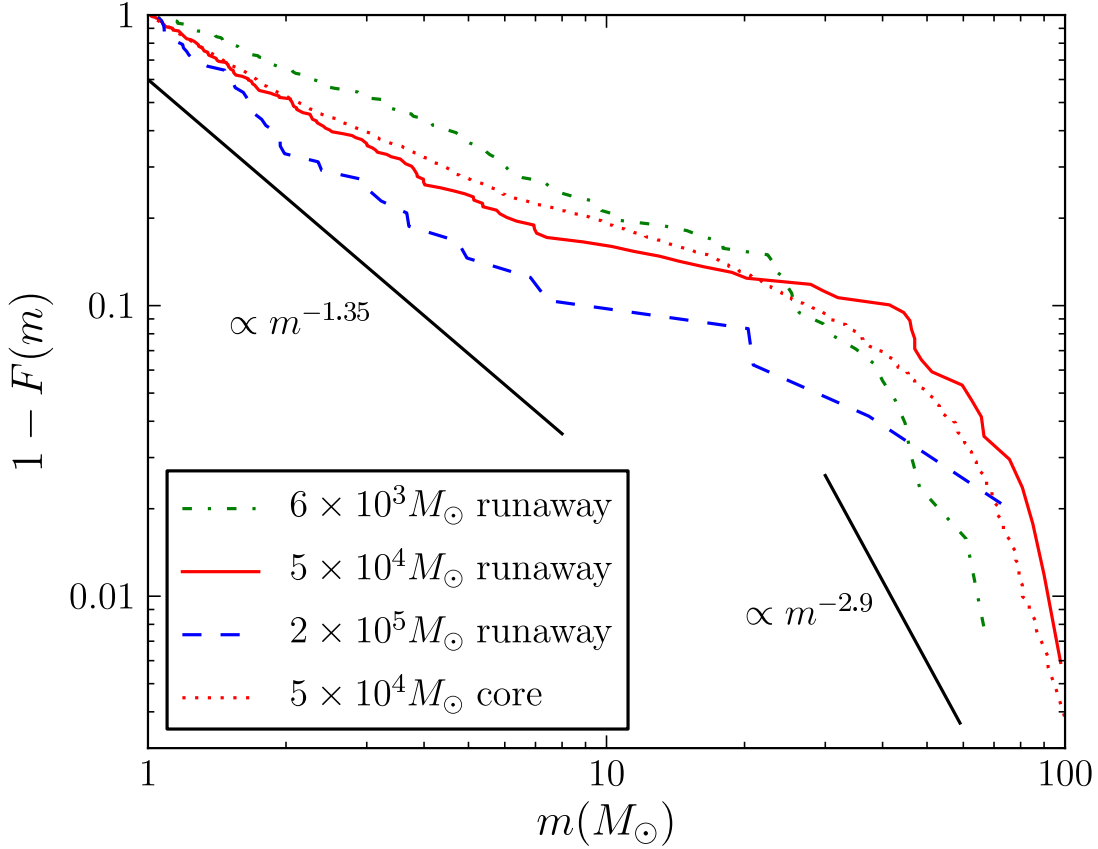,width=\columnwidth}
\caption{The cumulative mass functions ($1-F(m)$) of runaway and core
  ($r<0.1$pc) stellar populations. The solid black lines indicates the
  Salpeter initial mass function (left) and the observed mass function
  for massive field stars and runaways in the small Magellanic cloud
  \cite{2011arXiv1109.6655L} (right black curve).  The high-mass end
  of the function of the runaway stars from the simulations are
  consistent with the core mass function (dotted red curve); all are
  strongly mass segregated. The break point of the power law for the
  runaway mass functions is around 10\,$M_{\odot}$ and agrees with the
  expectation from the energy equipartition
  \cite{2007MNRAS.378L..29P,2008ApJ...686.1082F}.  The green
  dash-dotted, red solid and blue dashed curves give the results for
  simulated star cluster with a total mass of 6\,300\,\Msun\, (model
  A), $5.1\times 10^4$\,\Msun\, (model C) and $2.0\times
  10^5$\,\Msun\, (model D) respectively (Tab.\,S1).  }
\label{fig:mf}
\end{center}
\end{figure}

In our simulations, the large mass and relatively wide orbits of the
binaries produced during the gravothermal collapse of a cluster core
completely dominate the cross section for stellar encounters
(Supplement\,\S\,A). The encounter probability then should depend only
weakly on the mass and size of the single stars in the cluster core.
As a result, the mass distribution of the ejected stars roughly follow
the mass function in the cluster core which should be relatively flat
compared to the global cluster mass function
\cite{2002ApJ...570..171F}, because equipartition of kinetic energy
has resulted in the massive stars to sink to the cluster center
(Supplement\,\S\,B).

In Fig.\,\ref{fig:mf} we compare the distribution of the masses of the
runaway stars with the mass function in the cluster core (see also
Supplement\, \S\,B).  The mass distribution of the runaway stars is
statistically indistinguishable from the core distribution.  For stars
of $m\aplt 10$\,\Msun\, the mass distribution of the runaways are
flatter than the Salpeter slope, whereas it is considerably steeper
for $m \apgt 30$\,\Msun.  The steep high-mass end is consistent with
the recently observed distribution of massive runaway stars and field
stars in the small Magellanic cloud \cite{2011arXiv1109.6655L}.

\begin{figure}[htbp]
\begin{center}
\psfig{figure=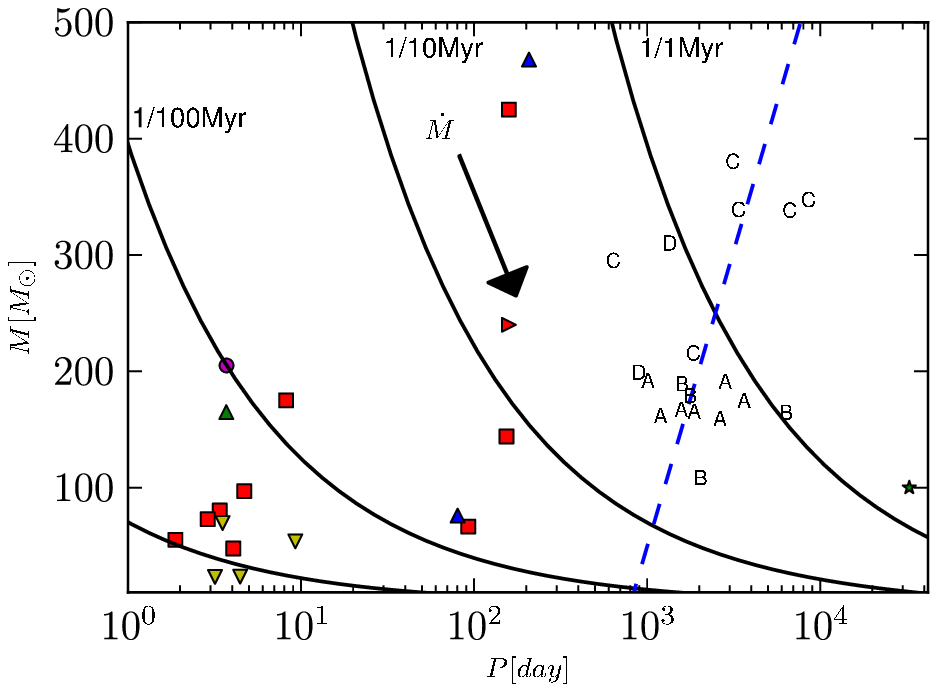,width=\columnwidth}%,angle=-90} 
\caption{The mass and orbital period of the binaries observed in
  nearby massive star clusters.  The observed binaries are from the
  star clusters R136 (red squares)
  \cite{2002ApJ...565..982M,2008MNRAS.389L..38S,2009MNRAS.395..823S,2011A&A...530L..10T},
  Trumpler~16 (blue up-pointed triangles) \cite{2008IAUS..250..119M},
  NGC3603 (magenta bullet) \cite{2008MNRAS.389L..38S}, Westerlund 1
  (yellow down-pointed triangles) \cite{2010arXiv1009.4709K},
  Westerlund 2 (green up-pointed triangle) \cite{1998A&AS..127..423P}
  and Trumpler~14 (green star) \cite{2009AJ....137.3358M}.  (For the
  binary R145 we have plotted also the latest preliminary parameters
  \cite{2011IAUS..272..497C} as the red right-pointed triangle.)  The
  simulation results are represented with the letters A, B, C and D
  for the respective models (Tab.\,S1).  The arrow (indicated with
  $\dot{M}$) gives the change in mass and orbital period of a binary
  as a result of adiabatic mass loss by a stellar wind. The blue
  diagonal dashed line (right) indicates the parameters for which the
  cross section for ejecting a single star with $v_{\rm ej} >30$km/s
  is maximal, adopting an equal-mass interaction (Supplement\,\S\,A).
  The solid black curves give the encounter rate (as indicated) for
  parameters in the core of R136 (adopting a core radius of 0.1\,pc).
\label{fig:ClusterBinaries}}
\end{center}
\end{figure}

The binaries from the $N$-body simulations populate the area around
the dashed curve in Fig.\,\ref{fig:ClusterBinaries}.  These binaries
came from a longer orbital period and lower mass to reach their
minimum orbital period mediated by encounters and collisions
(Supplement \S\,A).  Binaries to the right of this curve are in the
process of dynamical hardening by ejecting stars from the cluster;
they most efficiently produce runaways.  When they move away from the
dashed blue curve their encounter frequency drops and stellar mass
loss in the winds of the stars start to dominate their evolution
towards lower masses and longer orbital periods
(Fig.\,\ref{fig:ClusterBinaries}).

The short $\aplt 10$\,day period binaries observed inside young star
clusters like R136, must be primordial and probably never experienced
a dynamical encounter with another cluster member
(Fig.\,\ref{fig:ClusterBinaries}).  From a dynamical perspective these
binaries can be considered as single stars \cite{2009PASJ...61..721T},
because their interaction cross section is very small (Fig.\,S1). A
strong interaction with such a tight binary tends to cause a collision
between two or all three stars
\cite{2004MNRAS.352....1F,2011MNRAS.410..304G}; $\sim 80$\,\% of the
collision products remain in the cluster, the others escape
(Supplement \S\,A).

The relatively wide ($P \simeq 90$\,year) binary HD93129AB
\cite{2004AJ....128..323N,2009AJ....137.3358M} (green star in
Fig.\,\ref{fig:ClusterBinaries}) probably formed during the early
collapse of Trumpler~14
\cite{2010A&A...515A..26S,2005hst..prop10602M}. With the observed mass
and density of Trumpler~14 \cite{2008NewA...13..508O} this binary
would be about 2,700\,kT, consistent with being formed dynamically
during core collapse and currently in the process of hardening, but
insufficiently hard to eject itself from the cluster.  We expect that
HD93129AB has ejected $\sim 16$ stars, a few of which with
$m>8$\,\Msun.

The outskirts of the star cluster R136 also exhibits evidence of its
violent dynamical history.  The rapidly rotating 150\,\Msun\, star
VFTS~682 \cite{2011A&A...530L..14B} and the single 90\,\Msun\, star
30Dor~016 \cite{2010ApJ...715L..74E}, ware recently demonstrated to
once have belonged to the cluster.  The proper motion and projected
distance from R136 indicates that VFTS~682 and 30Dor~016 were ejected
29[pc]/30[km/s] $\simeq 1.0$\,Myr and 120[pc]/85[km/s] $\simeq
1.4$\,Myr ago, respectively, shortly after the birth of the $\sim
2$\,Myr old cluster. By this time R136 was much too young to have
experienced a supernova, challenging the favorite explanation for the
origin of OB runaway stars by the supernova explosion of a stellar
companion \cite{1961BAN....15..265B}.

The binary R145 is sufficiently hard to be responsible for having
ejected VFTS~682 and 30Dor~016 from R136, but so massive ($M_b \apgt
400$\,\Msun\, \cite{2009MNRAS.395..823S,Note_R145AsMassiveBinary})
that conservation of linear momentum would not impart a high enough
velocity kick to its center of mass to eject itself from R136.

Diagonally opposite R145 with respect to the center of R136 is the
x-ray source 16, which is an ACIS \cite{Note_ACIS} source with
a visual as well as a 2MASS counterpart \cite{2006AJ....131.2164T}.
Conservation of linear momentum may have caused both objects to be
ejected in opposite directions with respect to R136.  Based on the
respective projected distances of these two objects from the star
cluster we predict that R145 has ejected itself and the x-ray source
16 from R136.  The latter then exhibits a single $\sim 130$\,\Msun\,
star or a tight binary.

R145 and the other massive binary in Fig.\,\ref{fig:ClusterBinaries},
WR25 (HD93162) in the Carina region \cite{2007ASPC..367..159V}, are
representative of the population formed dynamically during the core
collapse of the nearby parent cluster, and flung out by a dynamical
interaction.  Neither are currently in the central portion of the
nearby clusters, R136 and Trumpler~16, respectively.  Both clusters
must therefore be in a later stage of core collapse where the binary
burning has stopped at the cost of the ejection of a rich population
of runaways, whereas the core of Trumpler~14 collapsed only recently.

\begin{figure}[htbp!]
\psfig{figure=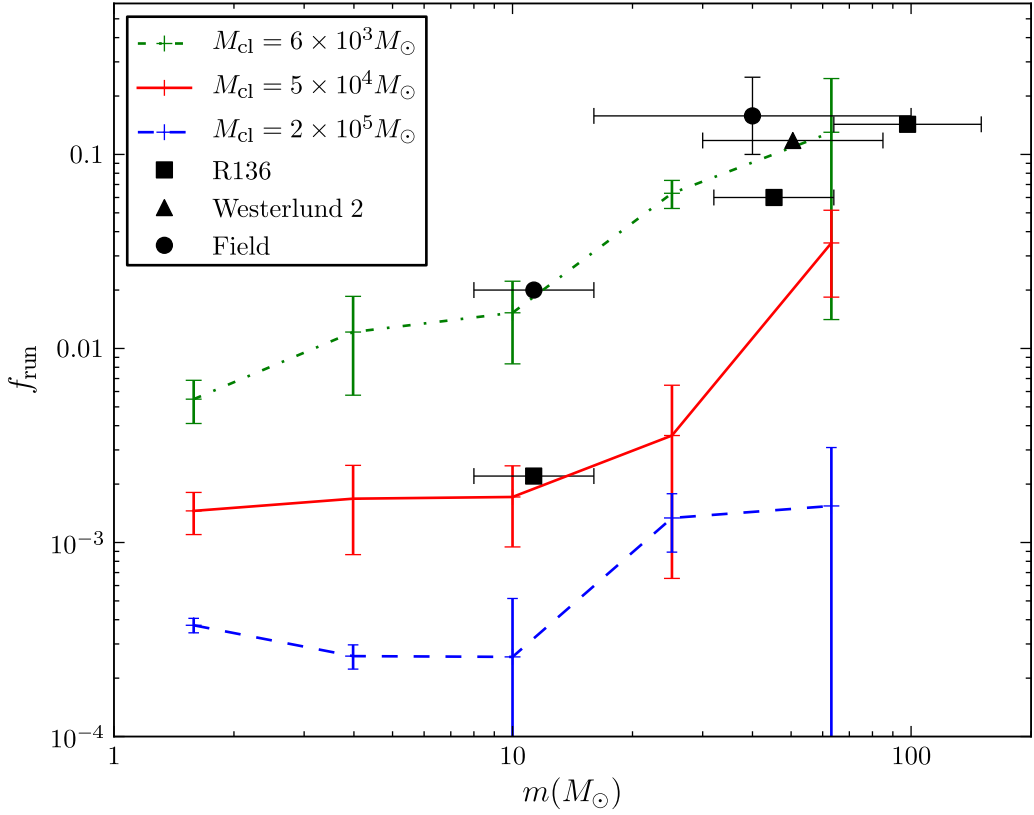,width=\columnwidth}%,angle=-90}
\caption[]{The relative fraction of runaways $f_{\rm run}$ as a
  function of stellar mass $m$.  $f_{\rm run}$ represents the number
  of runaway stars with mass $m$ divided by the total number of stars
  in the cluster with the same mass.  The squares, up-pointed triangle
  and bullets give the fraction of runaways found to be associated
  with the star clusters R136, Westerlund~2 and the Galactic field
  population, respectively. The normalization for the clusters is
  realized by counting the number of cluster members of that mass.
  The line styles are as in Fig.\,\ref{fig:mf}.
\label{Fig:EscapeFraction}
}
\end{figure}

Because each cluster generates one runaway-producing binary during
core collapse the relative fraction of runaways $f_{\rm run}$ is
inversely proportional to the mass of the cluster
(Fig.\,\ref{Fig:EscapeFraction}). A more complete census of OB
runaways can be obtained in the disk of the Milky Way in which more
than 634 runaway stars have been identified
\cite{1986ApJS...61..419G}.  The mass function of these runaways is
consistent with that of our simulations of young core-collapsed star
clusters of 6300\,\Msun\, (Fig.\,\ref{Fig:EscapeFraction}).

The majority of the galactic OB runaways seem to originate from star
clusters that experienced core collapse within the first 1\,Myr of
their existence.  The relative fraction of field runaways
(Fig.\,\ref{Fig:EscapeFraction}), is somewhat higher than in R136. For
stars with a mass $\apgt 8$\,\Msun\, this fraction is consistent with
the stellar ejecta produced in relatively low mass $\sim
6300$\,\Msun\, star clusters that experience core collapse before they
turn 1\,Myr old, whereas the R136 results matches the simulations with
a mass of $5\times 10^4$\,\Msun.  Such star clusters can experience
core collapse within about a million years if they are born highly
concentrated.

\section*{Supporting Online Material}
SOM Text \\
Figs. S1 to S4\\
Table S1\\
References (39 --- 57)\\

\appendix{}

In the main paper we discuss the ejection of stars through a
gravitational interaction from a binary system.  Here we present the
more technical part of our discussion. In particular the discussion on
three-body interactions in the cluster core and the macroscopic
dynamics of the star cluster.  This combination of microscopic
gravitational dynamics together with the global evolution of the star
cluster allows us to quantify the processes that lead to the ejection
of massive runaway stars and at the same time qualify these processes
by means of global structure evolution calculations of the stellar
clusters of interest. Our focus is generally on nearby young and
massive clusters. From an observational point of view these clusters
are well studied, and form a pivotal role in the understanding of the
formation of star clusters.  We aim our analysis towards populations
of young and massive clusters from Tab.\,2 of
\cite{2010ARA&A..48..431P}.

The cluster R136 is of particular interest to us, because it is well
studied and a rich population of runaways was recently identified to
roam the neighborhood.  With a mass of $M \simeq 60,000$\,\Msun, a
core radius of $r_{\rm core} \aplt 0.10$\,pc
\cite{2009ApJ...707.1347A,2010ARA&A..48..431P} and a virial radius
$r_{\rm vir} \simeq 2.89$\,pc \cite{2003MNRAS.338...85M} the density
profile of R136 fits a King \cite{1966AJ.....71...64K} model with a
dimension-less central potential $W_0 \simeq 9.9$.  As a consequence
the density in the cluster core $\rho_{\rm core} \apgt
47,000$\,\Msun/pc$^3$, which is quite typical for a core collapsed
star cluster \cite{1996yCat.7195....0H} (see also
Fig.\,\ref{Fig:densityEvolution}). Inside this cluster close
encounters between stars occur quite frequently. Most common are the
encounters between a single star and a binary. In
\S\,\ref{Sect:scatter3} we discuss the results of a series of
numerical scattering experiments between a binary and a single
star. We attempted to match the initial conditions for the scattering
experiments to mimic the environment in the core of R136 and in
particular the runaway 30Dor~016.  In \S\,\ref{Sect:NBody} we report
on N-body simulations of star clusters which bracket the range of
initial masses of observed star clusters, and the cluster size is
chosen such that they experience core collapse within about one
million years.

\section{Numerical three-body scattering experiments}\label{Sect:scatter3}

Here we specifically address the relation between the interaction
cross section and the binary parameters in order to understand the
orbital characteristics of the binaries that are likely to eject
massive stars from a star cluster.  The conclusions drawn from the
results presented in this supplement support the claim in the main
paper that relatively wide ($P \simeq 1000$\,days) and massive
binaries are most efficient in ejecting stars from the cluster,
whereas tight ($\aplt 10$ day) binaries are unlikely to interact and
therefore likely to be primordial. If binaries with orbital parameters
similar to the observed population of short period binaries in young
and massive star cluster interact dynamically they either cause a
collision between stars or eject themselves from the cluster.

We determine the interaction cross section by performing a series of
three-body scattering simulations using the {\tt sigma3} package
within the starlab software environment \cite{2001MNRAS.321..199P}.
In {\tt sigma3} cross sections are determined by executing a number of
three-body simulations in which a single star is injected into a
specific binary.  The encounter is initialized using the masses and
radii of all the three stars, and the relative velocity between the
binary center of mass and the intruder. The direction from which the
intruding star approaches the binary and the binary eccentricity are
selected randomly. The latter from the thermal distribution ranging
from a circular orbit to a maximum eccentricity such that the
stars do not touch at pericenter.  The radii of the stars are
calculated adopting zero-age main-sequence stars with solar
composition using MESA\,\cite{2004PASP..116..699P} called from the
AMUSE framework \cite{2009NewA...14..369P}. The orbital phase and the
direction from which the intruder approaches the binary are properly
randomized \cite{1983ApJ...268..319H}.  The three-body cross section
is subsequently determined, as described in
\cite{1996ApJ...467..348M}.  All calculations are run with a relative
energy error $\Delta E/E < 10^{-5}$.

\begin{figure}[htbp!]
\psfig{figure=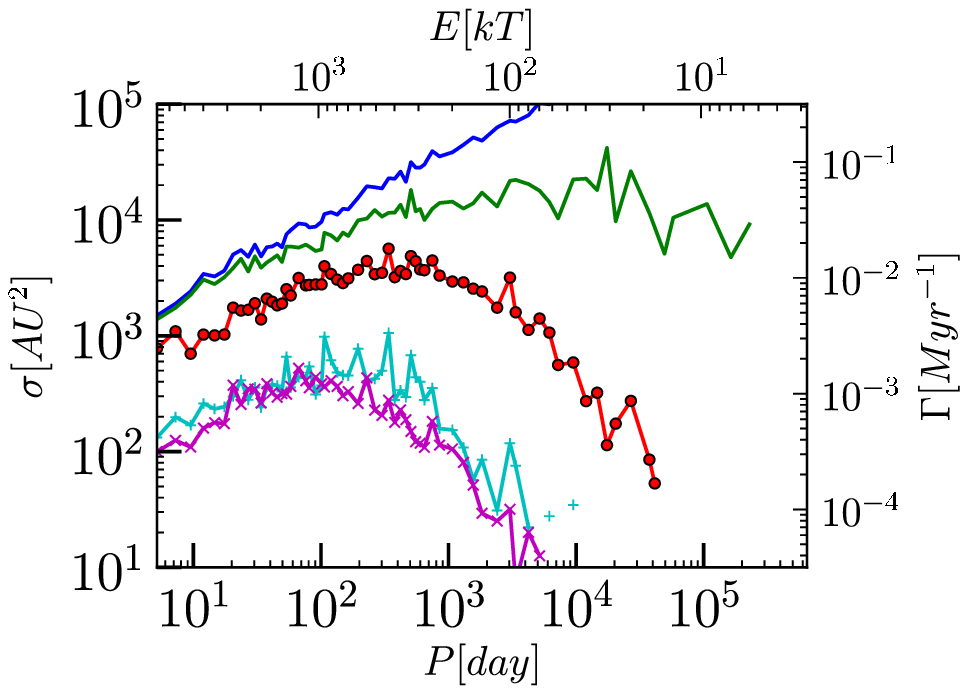,width=0.5\columnwidth}%,angle=-90}
~\psfig{figure=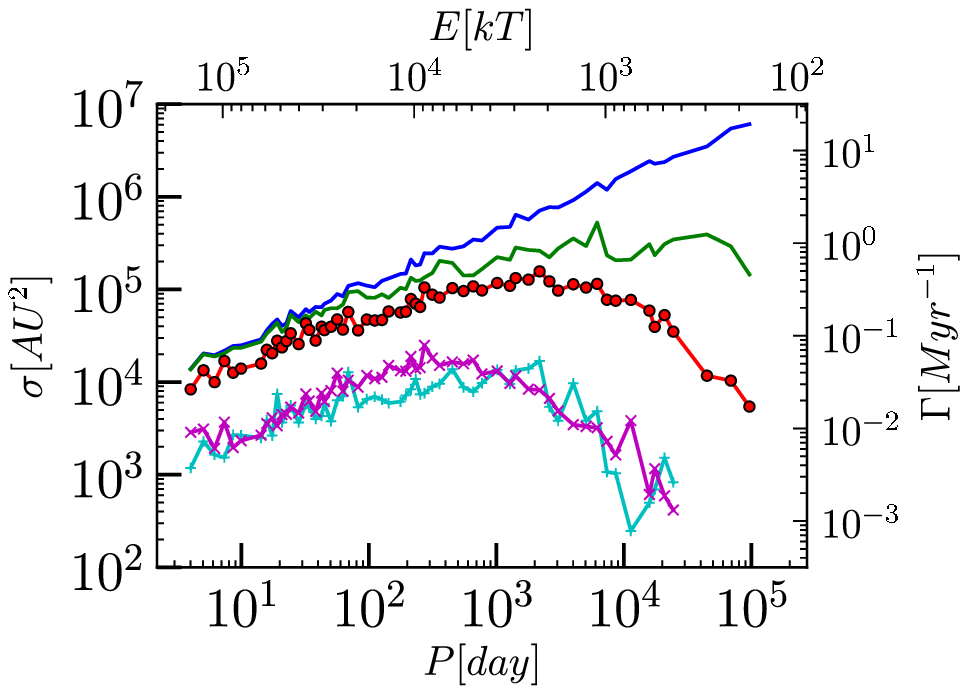,width=0.5\columnwidth}%,angle=-90}
\caption[]{The cross section for an encounter between a single star
  and a binary as a function of the orbital period. The left panel
  gives the results for single star and binary stars of 16\,\Msun,
  whereas the right-hand panel is for 90\,\Msun\, stars.  The top
  (blue) curve in both panels indicates the cross section for an
  exchange interaction or a collision between two of the stars or all
  three stars.  The green curve gives the cross section for
  collisions, which is a subset of the top (blue) curve.  The red
  curve with bullets gives the cross section for encounters that lead
  to the ejection of one of the stars to a velocity $>30$\,km/s.  The
  bottom two curves give the cross section for encounters that lead to
  the ejection of a merger product (magenta) or of a binary (light
  blue).  Along the right-hand vertical axis the cross section is
  expressed in terms of the encounter rate $\Gamma$ for which we
  adopted cluster parameters similar to R136. Along the top horizontal
  axis we express the encountering binary in terms of binding energy
  in units of kT adopting a mean mass $\langle m\rangle = 3.1$\,\Msun,
  which is consistent with the average stellar mass in our N-body
  simulations A, C and D (see \S\,\ref{Sect:NBody} and
  Tab.\,\ref{tb:model}).}
\label{Fig:ScatteringCrossection}
\end{figure}

In Fig.\,\ref{Fig:ScatteringCrossection} we present the results of a
series of scattering experiments in which we launch a main-sequence
star into a binary consisting of two stars of the same mass as the
incoming star. The left panel in Fig.\,\ref{Fig:ScatteringCrossection}
is calculated using 16\,\Msun\, stars with a radius of 4.9\,\Rsun\,
whereas the right panel is calculated using 90\,\Msun\, stars with a
radius of 12.8\,\Rsun.  We deliberately selected the single stars and
the binary components to have the same mass to make the interpretation
easier, and to prevent selection effects due to the sampling of the
other binary parameters.  We performed additional simulations using
stellar masses of 150\,\Msun\, 200\,\Msun\, and 400\,\Msun\, The blue
dashed curve in Fig.1 of the main paper is a least squares fit to the
stellar mass and the maximum cross section for ejecting a star with a
velocity of $>30$\,km/s, which we calculated by means of scattering
experiments.  This curve is consistent with a constant cross section
through mass and orbital period.

The bottom x-axis and left-hand y-axis in
Fig.\,\ref{Fig:ScatteringCrossection} give initial orbital period of
the target binary (in days) and the cross section (in units of ${\rm
  AU}^2$) resulting from the scattering experiments.  Each bullet
point in Fig.\,\ref{Fig:ScatteringCrossection} results from a
encounter density of 3000 per surface area of the binary, totaling to
about $10^5$ scattering experiments. The various curves are linear
interpolations between two bracketing points.  The top (blue) curve in
Fig.\,\ref{Fig:ScatteringCrossection} gives the total cross section
for exchanges and collisions (between two or all three stars).  The
green curve presents the cross section for collisions only.  After the
encounter is resolved, we record the velocity of the escaping
star. The red bullets give the cross section for an encounter which
leads to an escaper with an ejection velocity of at least $30$\,km/s,
which is sufficient to escape from a cluster like R136 and to be
considered a runaway star.  The blue (``+'' signs) and magenta
(``$\times$'' signs) curves present the cross section for a merger
product and a binary with an ejection velocity $>30$\,km/s,
respectively.

Along the top x-axis of Fig.\,\ref{Fig:ScatteringCrossection} we
present the hardness of the binary (in terms of kT, where kT is the
thermodynamic unit of kinetic energy in the cluster with ${3\over2}
NkT$ being the total binding energy for a cluster of $N$ stars) with
an orbital separation consistent with the bottom x-axis.  The binding
energy of the binary is calculated adopting cluster parameters
consistent for R136.  Along the right-hand y-axis we express the cross
section $\sigma$ (left hand y-axis) in terms of the encounter rate
between the binary and any other star $\Gamma$ (in units of
Myr$^{-1}$), which we calculated by adopting a central velocity
dispersion of $v_{\rm disp} = 10$\,km/s.

A binary of two 16\,\Msun\, stars (the minimum for a main-sequence
spectral type O star) with a $\sim 31$\,year orbital period can
already eject an incoming 16\,\Msun\, star with a space velocity of
$v_{\rm ej} \simeq 30$\,km/s. In the core of a typical young and dense
star cluster a binary with these parameters exhibits a binding energy
of several 10\,kT.  Although such a binary is wide, the cross section
for producing a runaway in an interaction is relatively small, but
rapidly increases with decreasing orbital period, until a maximum is
reached at a period of a few years (see
Fig.\,\ref{Fig:ScatteringCrossection}).  The maximum cross section for
this binary for producing a runaway with $v_{\rm ej} > 30$\,km/s peaks
at an orbital period of 800\,days ($\sim 300$\,kT) with an encounter
rate of $\Gamma \simeq 0.01$\,Myr$^{-1}$.  For higher mass stars
(90\,\Msun) the peak shifts to 2100\,days for an encounter rate of
$\Gamma \simeq 0.3$\,Myr$^{-1}$.  These rates are consistent with our
global estimate in the main paper for the orbital separation and mass
of a binary that most efficiently ejects an incoming star from the
cluster.

For larger orbital period (softer binaries) the cross section for
producing a runaway drops.  For short orbital period ($P \aplt
100$\,day) the cross section is dominated by collisions, which is
consistent with the result presented by \cite{2011MNRAS.410..304G},
although they adopted $a = 55$\,\Rsun\, with stellar masses comparable
to those adopted in the bottom panel of
Fig.\,\ref{Fig:ScatteringCrossection}.

The cross section for a collision exceeds the cross section for
producing a runaway over the entire range of orbital separations, and
this difference becomes more prominent for wider binaries. The
majority of these collision products, remain in the cluster, but a
small fraction of about 1/10 escape. The fraction of merger products
among the escaping stars hardly depends on the orbital separation, at
least up to $P \apgt 100$\,day (compare the red bulleted curve with
the lower magenta curve in Fig.\,\ref{Fig:ScatteringCrossection}).
The fraction of collision products among the runaways is then expected
to be rather constant, in particular because this trend appears to be
rather insensitive to the adopted initial conditions. For each five
escaping normal stars we then expect about double that number of
mergers to remain in the cluster and one merger product to escape as a
runaway.

The cross section for escaping wide binaries is comparable to that for
escaping merger products. This cross section does not include
$<10$\,day primordial binaries which dynamically behave similar
fashion as single stars \cite{2009PASJ...61..721T}.

\section{Simulations of massive young star clusters}\label{Sect:NBody}

To quantify the arguments concerning the orbital characteristics of
the binaries that form during core collapse, their efficiency of
ejecting stars with a high velocity and the distribution of stellar
masses ejected, we performed a series of direct N-body simulations of
star clusters over a range of cluster mass and size.

\subsection{Method and Initial Conditions}

\begin{table}[htbp]
\begin{center}
\begin{tabular}{crcccccccccc}\hline
Name & $N_\star$ &$N_{\rm run}$& $M$ &$\langle m \rangle$ & $W_0$ & $r_{\rm vir}$ & $\sigma$& $t_{\rm rh}$& \multicolumn{2}{c}{$N_{\rm runaways}$}\\ 
      \hline
      &&&[$10^3{\rm M}_{\odot}$]&[${\rm M}_{\odot}$]&&[pc]&[km/s]&[Myr] &all & $m>8$\,\Msun \\ 
      \hline\hline
A &2048  &7& 6.4 & 3.1 & 2 & 0.11  & 11 & 0.37& $18.1\pm2.4$&$5.0\pm1.3$\\ 
B &8192  &4& 5.7 & 0.69& 2 & 0.11 & 11 & 1.2 & $42\pm13$&$3.8\pm1.9$\\  
C &16384 &6& 51  & 3.1 & 6 & 0.40   &  17 & 4.4 & $27.7\pm5.6$&$6.2\pm1.0$\\  
D &65536 &2& 200 & 3.1 & 8 & 1.2   & 19 & 44  & $24.4\pm2.0$&$3.0\pm1.0$\\  
%64k$^\star$ &1& 66 & 1.0& 6 & 0.15 & 1.7 & 17 & 15 & $146&7\\
\hline
\hline
\end{tabular}
\caption{Simulated cluster models\label{tb:model}. From left to right we
identify the model name, number of stars in the simulations, the
number of runs, the total mass of the cluster, the mean mass of the
stars, the dimension less depth of the central potential and the
virial radius. The subsequent columns give derived parameters, which
are the central velocity dispersion and the half-mass relaxation time.
The last two columns give the number of runaways (averaged over the
$N_{\rm run}$ performed runs) for all stars which acquire a velocity
$v_{\rm ej} >30$\,km/s, and for the stars with $m>8$\,\Msun\, that
escape.}
\end{center}
\end{table}

In our $N$-body simulations we varied the total cluster mass, its
size and the stellar mass function in order to study the
characteristics of the population of runaway stars.  In
Tab.\,\ref{tb:model} we present an overview of the simulations
performed.

%For each,
%Models     N_run_total          N_run_massive
%2k             18.1 \pm 2.4         5.0 \pm 1.3
%16k           27.7 \pm 5.6        6.2 \pm 1.0
%64k           24.4 \pm 2.0        3.0 \pm 1.0
%total          23.0 \pm 5.9         5.2 \pm 1.6

For all simulations we adopted the Salpeter initial mass function
\cite{1955ApJ...121..161S} between a lower limit $m_{\rm min}$, which
we varied between 0.2\,\Msun\, and 1\,\Msun\, and a fixed upper limit
of 100\,\Msun. The upper limit and the slope of the selected power-law
mass function resulted in most of the low-mass ($M < 10000$\,\Msun)
clusters to have a maximum mass that rarely exceeded 30\,\Msun.

In our simulated star clusters we varied the mass, the virial radius
and the concentration of the density profile, in such a way that the
central density remained roughly the same.  Each cluster was
initialized using a King \cite{1966AJ.....71...64K} model density
distribution with a dimension-less central potential, $W_0$, of 2 for
the low-mass clusters, 6 for the intermediate mass clusters and 8 for
the most massive clusters.  As a consequence, the central density for
all clusters $\sim 2\times 10^6 M_{\odot}$pc$^{-3}$.  

In the table (Tab.\,\ref{tb:model}) we present the initial conditions
and main results of our model simulations.  

\begin{figure}[htbp!]
\psfig{figure=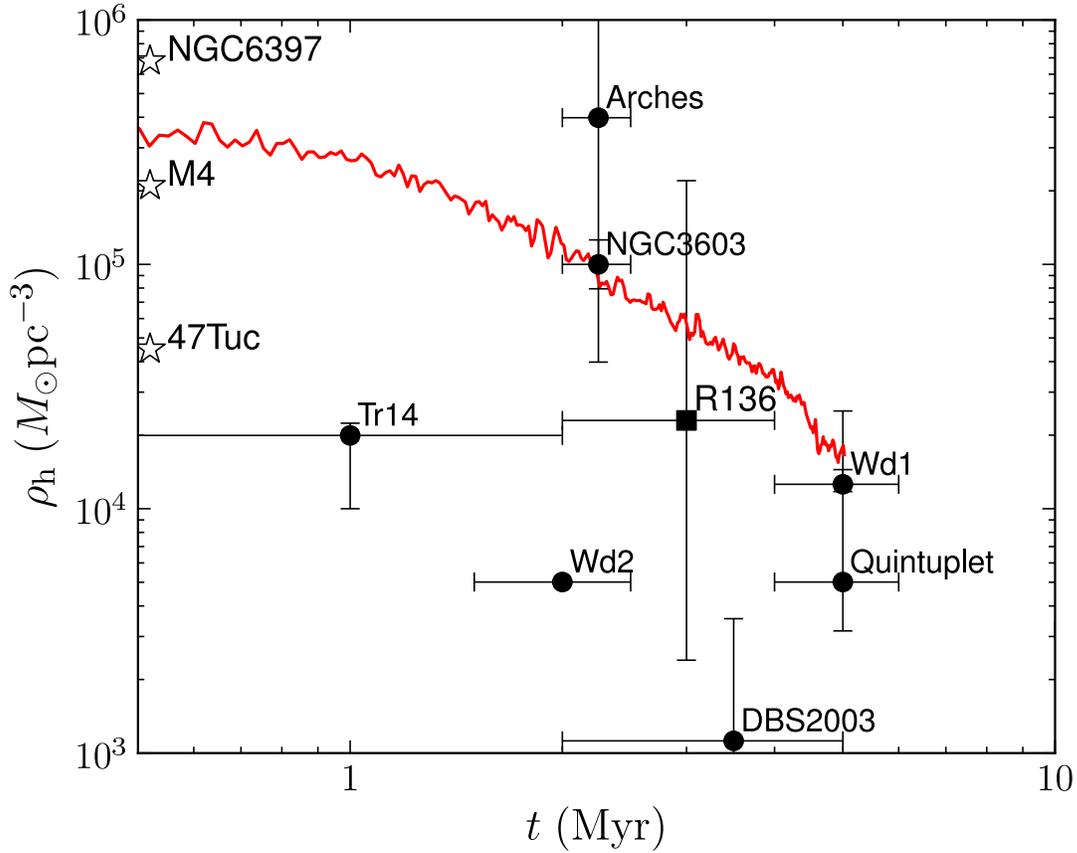,width=\columnwidth}%,angle=-90}
\caption[]{ The evolution of the half-mass density for simulations A,
  B and C (solid curve). Over-plotted using open stars symbols (to the
  left) are the initial density for the globular clusters
  NGC6397 \cite{2009MNRAS.395.1173G}, M4 \cite{2008MNRAS.389.1858H},
  and 47Tuc \cite{2011MNRAS.410.2698G}. The bullets with error bars
  are densities within the effective radius taken
  from \cite{2009A&A...498L..37P}. The half-mass density for R136 was
  taken from \cite{2009ApJ...707.1347A}.  }
\label{Fig:densityEvolution}
\end{figure}

In Fig.\,\ref{Fig:densityEvolution} we present the evolution of the
density within the half-mass radius for one of our simulations (C, see
Tab.\,\ref{tb:model}) and for comparison present the initial densities
for three globular clusters, and the observed densities of several
nearby young and massive clusters.  The half-mass densities for the
three globular clusters have been derived by iteratively performed
$N$-body simulations with primordial binaries and stellar evolution
until the final simulation results matches the observed star clusters
\cite{2008MNRAS.389.1858H,2009MNRAS.395.1173G,2011MNRAS.410.2698G}.
The initial density used in our model is bracketed by those derived
for the three globular clusters, and is quite consistent with the
observed values.

We run the simulations using a sixth-order Hermite predictor-corrector
integrator \cite{2008NewA...13..498N}, with a dimension-less accuracy
parameter of $\eta = 0.2$--0.3 \cite{2008NewA...13..498N}.  We
incorporated stellar collisions using the sticky-sphere approach for
which we adopted the stars to have a radius from solar composition
zero-age main-sequence stars \cite{2000MNRAS.315..543H}.  Initially
our simulations have no stars with a mass $>100$\,\Msun, but due to
collisions more massive stars started to appear shortly after core
collapse.  For stars of $m>100$\,\Msun, we adopted a mass loss by
stellar wind of $5.0\times 10^{-7} m$\,[yr$^{-1}$]
\cite{2009ApJ...695.1421F}, for lower mass stars we ignored the mass
loss.

The simulations were curried out using Cray XT4 in National
Astronomical Observatory of Japan with 64--512 cores for 16k and 64k
clusters and Little Green Machine in Leiden Observatory with 8 cores
for 2k clusters.

\subsection{Results from the $N$-body simulations}

We ran each simulation up to an age of 3\,Myr which is, for all the
initial conditions presented in Tab.\,\ref{tb:model}, well beyond the
moment of core collapse.  Core collapse in our models is a necessity
to develop a mass-segregated mass function and the dynamically active
bully binary in the core.  For each simulation we record the moment of
core collapse and the time of ejection, mass, and velocity of the
escaping stars.

In Fig.\,1 (of the main paper) we present the mass function of the
stars in the cluster core at an age of 3\,Myr. Except for the natural
run-to-run variations, the range of initial conditions did not show a
significant difference in the runaway or the core mass function. We
demonstrate this by presenting the core mass function for all runs
combined, but present the separate runaway mass function for the three
models A, C and D (see Tab.\,\ref{tb:model}). We compare this mass
function with the distribution of stellar masses that escaped the
cluster, with a Selpeter mass function and with the recently observed
distribution of massive runaway and field stars in the small
Magellanic cloud \cite{2011arXiv1109.6655L}.

The distribution of relatively low mass runaway stars $\aplt
8$\,\Msun\, is consistent with the Salpeter slope.  More massive stars
($m \apgt 8$\,\Msun) are significantly over represented among the
runaways and among the population in the cluster core.  The mass
function of the runaway stars with a mass $m \apgt 8$\,\Msun\, is
indistinguishable from the distribution of masses in the cluster
center.  Although the massive stars are over-represented among the
runaways and field stars (see Fig.\,1 of the main paper), the
high-mass end ($m\apgt 30$\,\Msun) distribution is considerably
steeper, as was also observed in the small Magellanic
cloud \cite{2011arXiv1109.6655L}. The consistency between the observed
and simulation results further support our finding that the majority
of massive stars are formed in relatively small (in size as well as in
mass) star clusters and that the ejection mechanism is dynamic.

As we discussed in the main paper, both distributions are statistical
indistinguishable because the cross-section for gravitational
interactions is dominated by the relatively wide binary formed
dynamically during the collapse of the cluster core; the distribution
of ejected stars depends only weakly on the mass of the single stars.
The mass function of the runaway stars depends slightly on the mass of
the cluster in the sense that the least massive clusters tend to have
a slightly steeper mass function. This is caused by the lower
probability of each individual low-mass cluster to contains a massive
star.

\subsection{Velocity distribution}

\begin{figure}[htbp]
\begin{center}
\psfig{figure=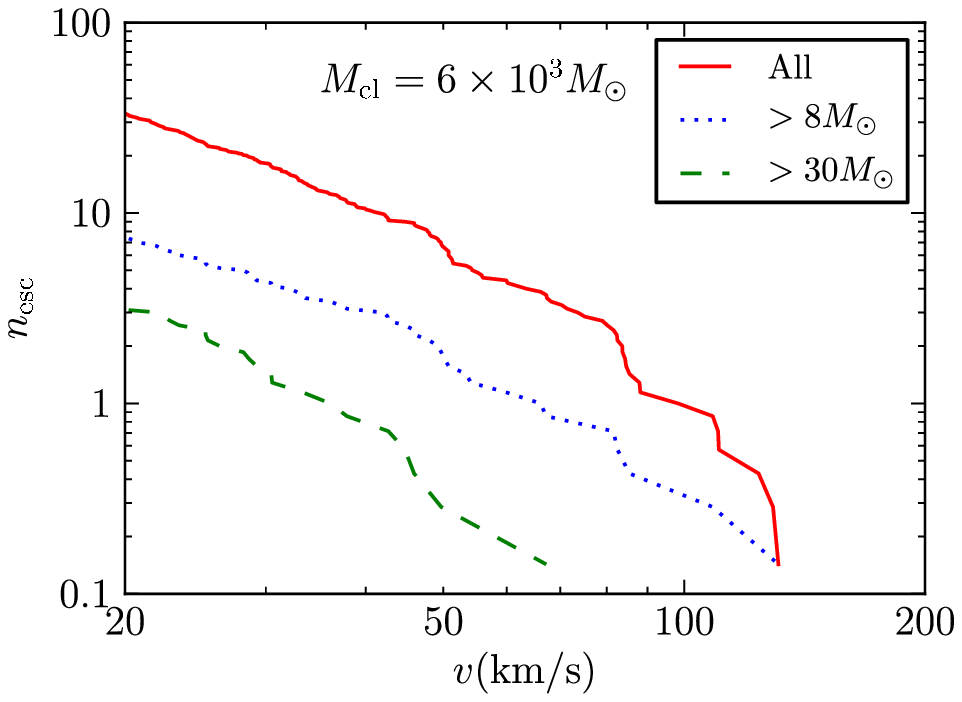,width=0.5\columnwidth}%,angle=-90} 
~\psfig{figure=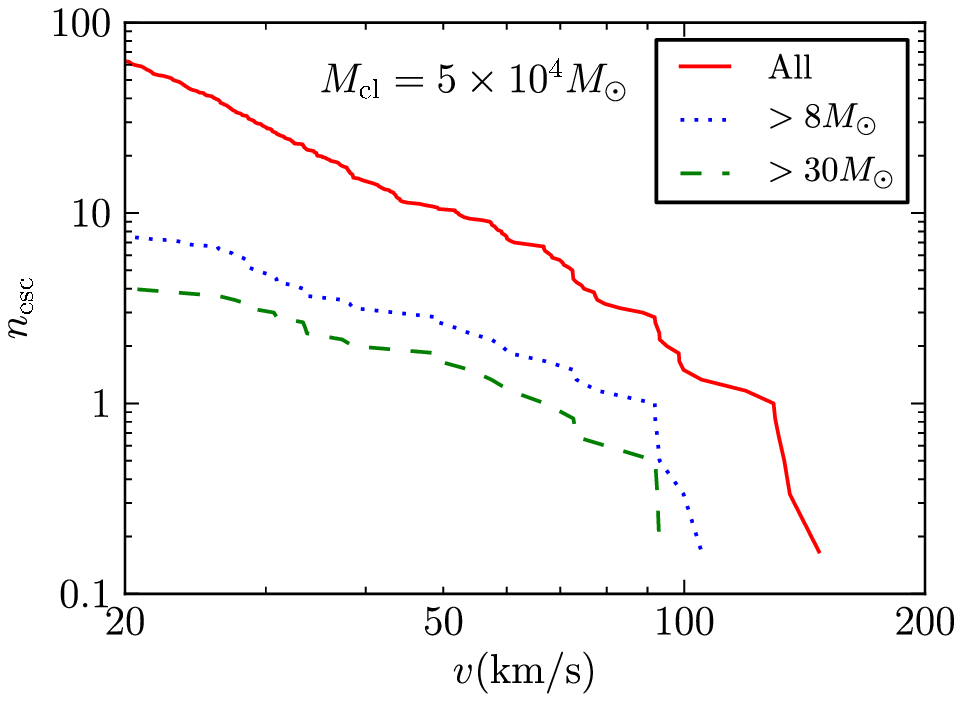,width=0.5\columnwidth}%,angle=-90} 
\caption{ The cumulative velocity distribution of runaway stars for our
  simulations with a minimum mass of 1\,\Msun\, and a total cluster
  mass of 6300\,\Msun\, (left, model A see also Tab.\,\ref{tb:model})
  and $5.1\times10^4$\,\Msun\, (right, model C). The top (solid red)
  curve gives the distribution of the velocities of all runaways with
  $v_{\rm ej} > 20$\,km/s. For the more massive clusters, the velocity
  distribution of the stars $m>8$\,\Msun\, (blue dotted curve) and for
  stars $m > 30$\,\Msun\, (green dashes) are somewhat shallower.  }
\label{fig:nv_esc}
\end{center}
\end{figure}

In figure \ref{fig:nv_esc} we present the velocity distribution of
runaway stars of our simulations with a minimum mass of 1\,\Msun\,
using 6300\,\Msun\, and $5.1\times10^4$\,\Msun. Opting for
0.2\,\Msun\, as a minimum mass to the initial mass function does not
affect the results significantly.

The velocity distribution of ejected stars is skewed towards the
cluster escape speed, but when the binary becomes harder and the
ejected star less massive, higher escape velocities become more
common. The number of runaway star depends on the number of binaries
formed during core collapse, and not on the mass of the cluster. The
distribution of stars ejected reflects the hardening process of a
single tailor-made binary that formed during core collapse. Because it
is a single binary that produces all runaways, the scattering process
in the cluster core drives the energy and rate of escapers.  The mass
distribution of the escapers is then consistent with the stellar
population in the cluster core.  In a single cluster, only one hard
binary forms, and the number of ejected stars reflects its hardening
process.  The number of runaways produced, their distribution in mass
and in velocity then does not depend on the mass of the cluster.

The characteristic runaway velocity of massive stars, however, does
depend on cluster mass, even though it is still the single binary
process that drives the runaways.  The more massive stars
$>8$\,\Msun\, tend to have a higher velocity in massive clusters than
in the lower mass clusters. This trend continues also for a more
massive selection of stars, of $>30$\,\Msun.  The lower velocities, on
average, of the more massive stars in the lower-mass clusters is
caused by the smaller proportion of massive stars. Ejecting a massive
star from a cluster is achieved most easily by a binary consisting of
massive stars. However, the lower mass clusters have a smaller
probability of containing a sufficiently large population of massive
stars to achieve this. The more massive cluster, therefore, are able
to eject massive stars with higher speed.  A similar effect was noted
in relation to stellar collisions \cite{2008MNRAS.384..376G}.

\begin{figure}[htbp!]
\begin{center}
\psfig{figure=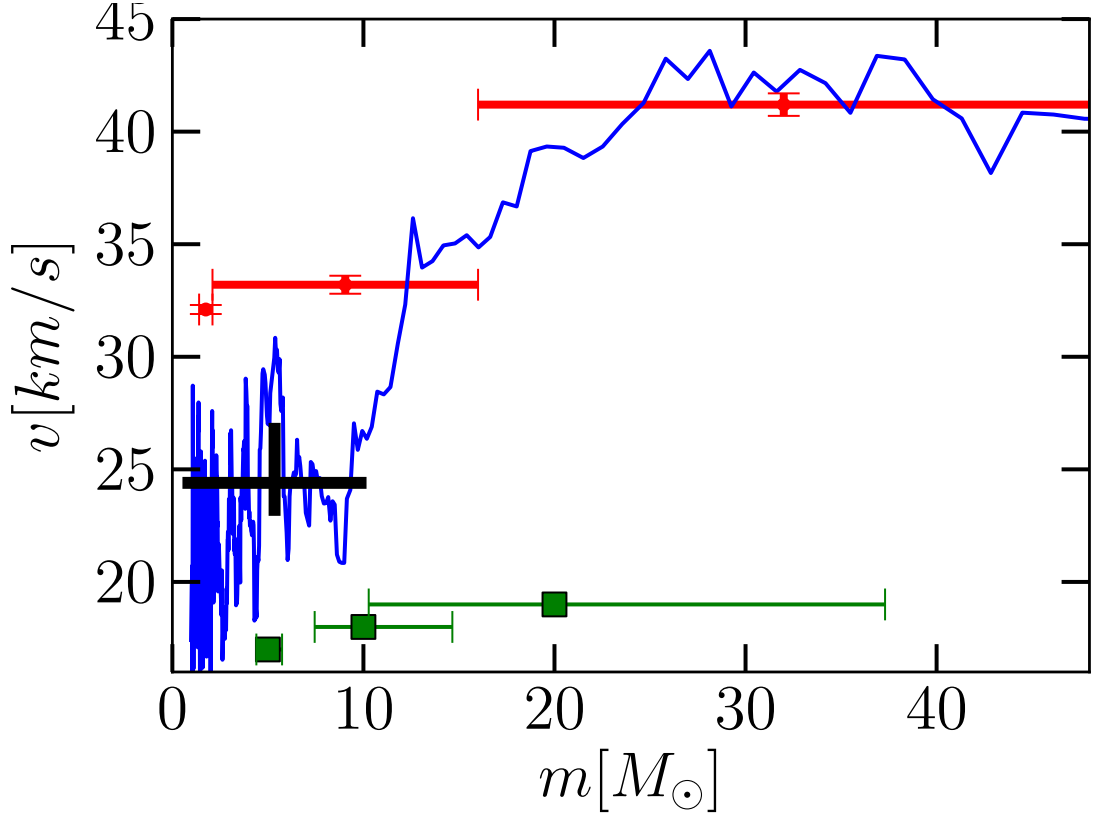,width=\columnwidth}%,angle=-90}
\caption[]{ The mean escape velocity as a function of mass. We
  recorded the masses and velocities of the stars that have escaped
  our simulated cluster (models C and D, see Supplement\,\S\,B) at an
  age of 3\,Myr. The blue curve is constructed by sorting all escaping
  stars in mass from least to most massive, and measure the means mass
  and their velocity in a co-moving window of 30 stars.  We plotted in
  green the size of the mass-window at the bottom of the panel for
  $m=5$\,\Msun, 10\,\Msun\, and 20\,\Msun\,(squares). The observed
  mean velocities of single main-sequence stars with spectral type A,
  B and O in the SMC \cite{2008MNRAS.386..826E} are presented in red
  (top horizontal bars), and the Galactic population of runaways from
  the Hipparcos database \cite{2011MNRAS.410..190T} is presented in
  black at 24.4\,km/s.  }
\label{Fig:EscapeVelocities}
\end{center}
\end{figure}

In Fig.\,\ref{Fig:EscapeVelocities} we present the space velocity as a
function of stellar mass.  The dark blue solid curve gives the mean
runaway velocity for the simulations C and D as a function of mass in
a co-moving bin of 30 stars. The horizontal bars (red) give the
observed mean velocity of stars in the small Magellanic cloud
\cite{2008MNRAS.386..826E}. For spectral type O stars ($m\apgt
16$\,\Msun) our simulations are quite consistent with the observed
velocities, although for later type spectral types our simulations
tends to somewhat under produce the velocities. Although using the
Hipparcos catalogue \cite{2011MNRAS.410..190T} have derived a mean
velocity of $24.4\pm1.2$\,km/s for runaways with a mean mass of
$5.4\pm4.5$\,\Msun. The average runaway velocity for stars with $m
\aplt 16$\,\Msun\, seem to be systematically higher in the SMC than
those observed in the Galaxy and our simulations.

\section*{Acknowledgments}
We thank B.\, Brandl, A.\, Brown, A.\, G\"urkan and H.\, Sana for
discussions, and the anonymous referees for their constructive
criticism.  This work was supported by Research Fellowships of the
Japan Society for the Promotion of Science (JSPS) for Young Scientists
and the Netherlands Research Council NWO (grants VICI [\#639.073.803],
AMUSE [\#614.061.608] and Little Green Machine) and by the Netherlands
Research School for Astronomy (NOVA).  The simulation software is
available at {\tt http://amusecode.org} and the data can be found at
{\tt http://initialconditions.org}.


\begin{thebibliography}{54}
\bibitem{Note_on_kT} 
We adopted kT as the thermodynamic unit of
  kinetic energy in the cluster; the binding energy for a cluster of
  $N$ stars is ${3\over2} N$kT.
\end{thebibliography}
\end{document}